# Microwave characterisation of $CaF_2$ at cryogenic temperatures using a dielectric resonator technique

Mohan V. Jacob[a,*], Janina E. Mazierska[a], Dimitri O. Ledenyov[a], Jerzy Krupka[b]

[a]*Electrical and Computer Engineering, School of Engineering, James Cook University, Townsville, QLD 4811, Australia*
[b]*Instytut Mikroelektroniki i Optoelektroniki Politechniki Warszawskiej, Koszykowa 75, 00-662 Warszawa, Poland*

**Abstract**

Properties of calcium fluoride ($CaF_2$) have been well researched at UV, visible and IR range of frequencies but not at microwave frequencies. In this work we report the loss tangent and the real part of relative permittivity $\varepsilon_r$ of $CaF_2$ measured in the temperature range 15–81 K and at frequency 29.25 GHz. The $\tan\delta$ and $\varepsilon_r$ were determined by measurements of the resonant frequency and the $Q_0$-factor of a $TE_{011}$ mode cylindrical copper cavity with superconducting plates containing the sample under test. The measured $\varepsilon_r$ of $CaF_2$ was found to change from 6.484 to 6.505, and the $\tan\delta$ from $3.1\times10^{-6}$ to $22.7\times10^{-6}$ when temperature was varied from 15 to 81 K. Due to the low losses $CaF_2$ can be useful in construction of high $Q$-factor microwave circuits and devices operating at cryogenic temperatures.

## 1. Introduction

Calcium fluoride crystals are used in many optical applications, including mirror substrates for UV laser systems, windows, lenses and prisms for ultraviolet, visible and infrared frequencies. $CaF_2$ is grown by the Stockbarger technique or the Brigdman method in diameter up to about 200 mm. Calcium Fluoride (VUV grade) crystals have the transmission range from 0.19 to 7.2 µm and low refractive index from about 1.35 to 1.51 through this range.[1] IR grade Calcium Fluoride is transparent up to 12 µm. Degradation due to moisture in the atmosphere is minimal, and polished surfaces may be expected to withstand several years exposure to normal atmospheric conditions. Due to its low refractive index, Calcium Fluoride can be used without an anti-reflective coating. The maximum temperature $CaF_2$ can tolerate is 800 °C in dry atmosphere. Low solubility and wide transmission makes this material useful for many applications, including mirror substrates for UV laser systems, windows, lenses and prisms for UV and IR applications.[1]

Due to the low relative permittivity and low losses, calcium fluoride can find applications in microwave planar circuits as a substrate material. Another possibility could be a hybrid high temperature superconductor (HTS)—silicon technology for microwave circuits as investigated in Refs. 2 and 3. As silicon atoms diffuse into HTS films during annealing at elevated temperatures resulting in deteriorating superconducting properties, $CaF_2$ was investigated for its usefulness to overcome this difficulty due to its chemical stability and structural and thermal compatibility with Si and GaAs.

Contrary to extensive data available for $CaF_2$ at optical frequencies, there is little data on microwave properties of this material, especially at cryogenic temperatures. In this paper we present results of precise measurements of the permittivity and loss tangent of $CaF_2$ at cryogenic temperatures from 15 to 81 K using the dielectric resonator technique. We have used the multifrequency Transmission Mode Q-Factor (TMQF) technique[4,5] for data processing to ensure high accuracy of calculated values of $\varepsilon_r$ and $\tan\delta$. Also the thermal expansion phenomenon of the material was taken into account in the calculations.

## 2. Dielectric resonator measurement method

The superconducting dielectric resonator technique is a modification of the metallic dielectric resonator[6,7] and has recently been used to characterise various low loss single crystal and polycrystalline dielectric materials at microwave frequencies.[8–10] The Hakki-Coleman version of the dielectric resonator we used for the measurements of CaF$_2$ is shown in Fig. 1. The resonator consisted of the copper cavity of diameter of 9.5 and 3 mm height with superconducting endplates. The CaF$_2$ sample was machined into a cylinder with the aspect ratio (diameter to height) equal to 1.67, with 3.1 mm height and 5.0 mm diameter.

* Corresponding author.

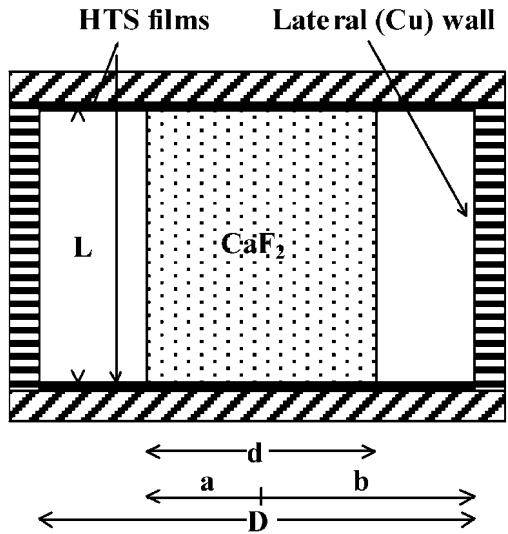

Fig. 1. Schematic of a TE$_{011}$ mode dielectric resonator.

The real part of relative permittivity $\varepsilon_r$ were determined from measurements of the resonant frequency of the resonator with the CaF$_2$ sample as the first root of the following transcendental equation[11] using software SUP12:[12]

$$k_{\rho 1} J_0(k_{\rho 1} b) F_1(b) + k_{\rho 2} J_1(k_{\rho 1} b) F_0(b) = 0 \quad (1)$$

where:

$$F_0(\rho) = I_0(k_{\rho 2}\rho) + K_0(k_{\rho 2}\rho) \frac{I_1(k_{\rho 2}a)}{K_1(k_{\rho 2}a)}$$

$$F_1(\rho) = -I_1(k_{\rho 2}\rho) + K_1(k_{\rho 2}\rho) \frac{I_1(k_{\rho 2}a)}{K_1(k_{\rho 2}a)}$$

$$k_{\rho 1}^2 = \frac{\omega^2 \varepsilon_r}{c^2} - k_z^2, \quad k_{\rho 2}^2 = k_z^2 - \frac{\omega^2}{c^2}, \quad k_z = \pi/L$$

and $\omega$ is the angular frequency ($2\pi f$), $c$ is velocity of light, $\varepsilon_o$ is free space permeability, $\varepsilon_r$ is real relative permittivity of the sample and $J_0$, $J_1$, $I_0$, $I_1$, $K_0$, $K_1$, denote corresponding Bessel and Hankel functions.

The loss tangent tan$\delta$ of CaF$_2$ was computed from the measured $Q_0$-factor of the resonator on the basis of the well known loss equation,[11] namely:

$$\tan\delta = \frac{1}{\rho_e} \left[ \frac{1}{Q_0} - \frac{R_{SS}}{A_S} - \frac{R_{SM}}{A_M} \right] \quad (2)$$

where $Q_0$ is the unloaded Q-factor of the entire resonant structure, $R_{SS}$ and $R_{SM}$ are the surface resistance of the superconducting and the metallic parts of the cavity respectively, $A_S$ and $A_M$ are the geometric factors of the superconducting part and metallic parts of the cavity and $\rho_e$ is the electric energy filling factor.

Geometric factors $A_S$, $A_M$, and $\rho_e$ to be used in (2) were computed using incremental frequency rules as follows:[11]

$$A_S = \frac{\omega^2 \mu_0}{4} / \frac{\partial \omega}{\partial L} \quad (3)$$

$$A_M = \frac{\omega^2 \mu_0}{2} / \frac{\partial \omega}{\partial a} \quad (4)$$

$$p_e = 2 \left| \frac{\partial \omega}{\partial \varepsilon_r} \right| \frac{\varepsilon_r}{\omega} \quad (5)$$

Computed values of the geometrical factors and the energy filling factors are given in Table 1.

Values of the surface resistance HTS endplates ($R_{SS}$) and copper walls ($R_{SM}$), necessary for Eq. (2) were measured in the same copper cavity but with the sapphire rod and results are given in Section 3.

## 3. Measurements of microwave properties of CaF$_2$

The measurement system we used for microwave characterisation of the calcium fluoride sample is shown in Ref. 10. The system consisted of Network Analyser (HP 8722C), closed cycle refrigerator (APD DE-204), temperature controller (LTC-10), vacuum Dewar, a PC and the Hakki-Coleman dielectric resonator in transmission mode. The CaF$_2$ sample was grown with the Brigdman method by Ref. 13.

### 3.1. Measurements of R$_{SS}$ and R$_{SM}$ of the Hakki-Coleman cavity

As mentioned in Section 2 the surface resistances $R_{SM}$ and $R_{SS}$ of the cavity needed to be measured first. To obtain precise values of $R_{SM}$ we have measured S-parameters ($S_{21}$, $S_{11}$, and $S_{22}$) around the resonance of the Hakki-Coleman resonator with the sapphire rod and copper cavity. The measured data sets were processed with the Transmission Mode Q-Factor Technique[3,4] to obtain the loaded $Q_L$-factor and coupling coefficients as mentioned in Section 1. The TMQF method accounts for noise, delay due to uncalibrated transmission lines and its frequency dependence, and crosstalk in measurement data and hence provides accurate values of $Q_L$ and the coupling coefficients $\beta_1$ and $\beta_2$. The unloaded $Q_0$-factor was subsequently calculated using the exact equation,[14]

$$Q_0 = Q_L(1 + \beta_1 + \beta_2) \quad (6)$$

Table 1
The geometrical factors and the energy filling factor of the Sapphire and CaF$_2$ resonators

| Dielectric rod | CaF$_2$ | Sapphire |
| --- | --- | --- |
| Frequency | 29.3 GHz | 24.65 GHz |
| $A_M$ | 22,029 | 22,319 |
| $A_S$ | 329.8 | 280.6 |
| $\rho_e$ | 0.96 | 0.97 |

Assuming loss tangent of the sapphire rod as $10^{-7}$, the surface resistance of copper $R_{SM}$ was calculated with Ref. 11 based on:

$$R_{SM} = A_M \left[ \frac{1}{Q_0} - \rho_e \tan\delta \right] \quad (7)$$

where $A_M$ is the metallic geometric factor for the copper cavity.

The surface resistance $R_{SS}$ was measured with the sapphire rod in the copper cavity with end walls comprising of a pair of high quality $YBa_2Cu_3O_7$ thin films and calculated with Ref. 12 based on:[11]

$$R_{SS} = A_S \left[ \frac{1}{Q_0} - \frac{R_{SM}}{A_M} - \rho_e \tan\delta \right] \quad (8)$$

Measured dependence of surface resistances, $R_{SS}$ and $R_{SM}$, with temperature at frequency of 24.6 GHz are presented in Fig. 2.

As measurements of calcium fluoride were performed at a frequency of approximately 29.25 GHz, not at 24.6 GHz, the measured values of $R_{SM}$ and $R_{SS}$ were scaled assuming the square root frequency dependence for $R_{SM}$ of copper and the frequency square law for $R_{SS}$ of superconducting endplates.

### 3.2. Measurements of microwave properties of calcium fluoride

The Hakki-Coleman resonator with HTS endplates containing the $CaF_2$ sample was cooled from room temperature to approximately 12 K, and the resonant frequency of 29.25 GHz was obtained. The $S_{21}$, $S_{11}$ and $S_{22}$ parameters data sets around the resonance were measured as a function of increasing temperature from 13 to 81 K, and the $Q_0$-factor and $f_{res}$ were calculated using the TMQF technique and Eq. (6) as before. The real relative permittivity $\varepsilon_r$ of the calcium fluoride sample was calculated from the measured resonant frequency using Eq. (1). Variation of dimensions of the $CaF_2$ sample with temperature as shown in Fig. 3 were taken into consideration in the computations of $\varepsilon_r$ using the temperature dependence of the linear thermal expansion coefficient after Ref. 15 as given in Fig. 4.

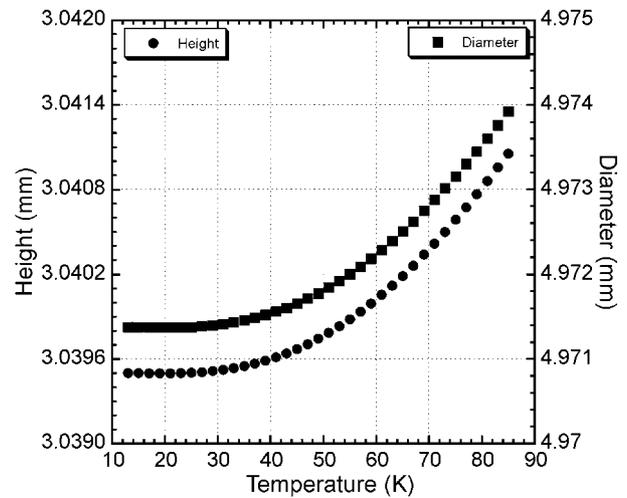

Fig. 3. Dimensions of the $CaF_2$ sample at cryogenic temperatures.

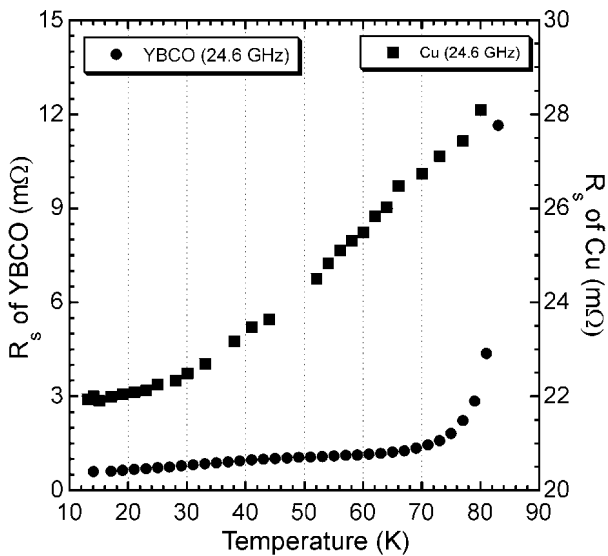

Fig. 2. Measured surface resistances of $YBa_2Cu_3O_{7-\delta}$ films and copper versus temperature.

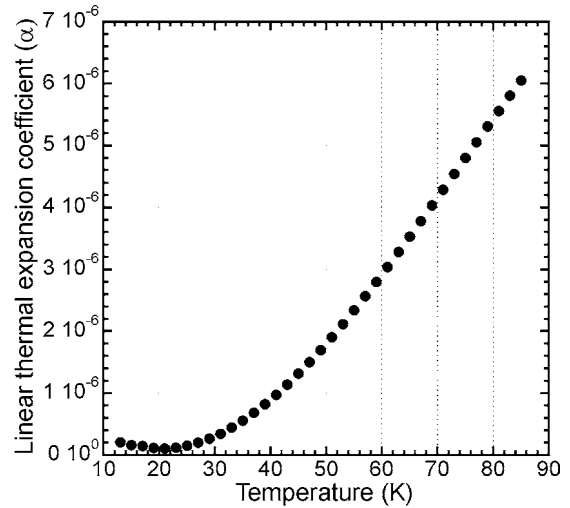

Fig. 4. Temperature coefficient of $CaF_2$ based on Ref. 15.

The real part of relative permittivity $\varepsilon_r$ of the CaF$_2$ sample measured at temperatures from 15 to 81 K is shown in Fig. 5. The $\varepsilon_r$ exhibited the magnitude of approximately 6.5 and increased with the temperature by approximately 0.33%; from 6.483 to 6.505.

The loss tangent of the CaF$_2$ sample was calculated using (2) from the measured unloaded $Q_0$-factor. The measured temperature dependence of tan$\delta$ is shown in Fig. 6. The loss tangent showed an increase of 86% in the temperature range from 15 to 81 K; at temperatures of 15 and 81 K, the measured tan$\delta$ of CaF$_2$ was $3.1 \times 10^{-6}$ and $2.27 \times 10^{-5}$, respectively. Using a linear scaling, the calculated loss tangent of CaF$_2$ at temperature of 15 K and frequency of 10 GHz is only $1.05 \times 10^{-6}$. Our results show that CaF$_2$ exhibits very low losses at microwave frequencies and cryogenic temperatures, comparable to losses in Teflon.[9]

## 4. Error analysis of measured parameters $\varepsilon_r$ and tan$\delta$ of CaF$_2$

The accuracy of the measurement of the real part of permittivity using the dielectric resonator depends on the precision of the measurements of the resonant frequency and uncertainty in dimensions of the dielectric sample. We measured $f_{res}$ with a resolution of 1 Hz using the Network Analyser HP 8722C. Hence to assess uncertainty in $\varepsilon_r$ measurements the error analysis was performed assuming the uncertainty in the dimensions of the sample of 0.2 and 0.5% using the software SUP12.[11] Results of the error analysis are presented in Fig. 7 and show that the relative error $\Delta_r \varepsilon_r$ is approximately twice the uncertainty in dimensions.

The uncertainty in the loss tangent measurements is caused by the uncertainty in the measured unloaded $Q_0$-factor values, $R_{SS}$, $R_{SM}$ and geometrical factors. The Most Probable Error (MPE) in tan$\delta$ of CaF$_2$ can be expressed after Ref. 10 as:

$$\nabla_r \tan\delta = \left[ \left| \frac{\Delta\rho_e}{\rho_e} \right|^2 + \left| \left( \frac{-1}{Q_0 \rho_e \tan\delta} \right) \frac{\Delta Q_0}{Q_0} \right|^2 \right.$$
$$+ \left( \frac{R_{SS}}{A_S \rho_e \tan\delta} \left( \left| \frac{\Delta R_{SS}}{R_{SS}} \right| + \left| \frac{\Delta A_S}{A_S} \right| \right) \right)^2 \quad (9)$$
$$\left. + \left( \frac{R_{SM}}{A_M \rho_e \tan\delta} \left( \left| \frac{\Delta R_{SM}}{R_{SM}} \right| + \left| \frac{\Delta A_M}{A_M} \right| \right) \right)^2 \right]^{1/2}$$

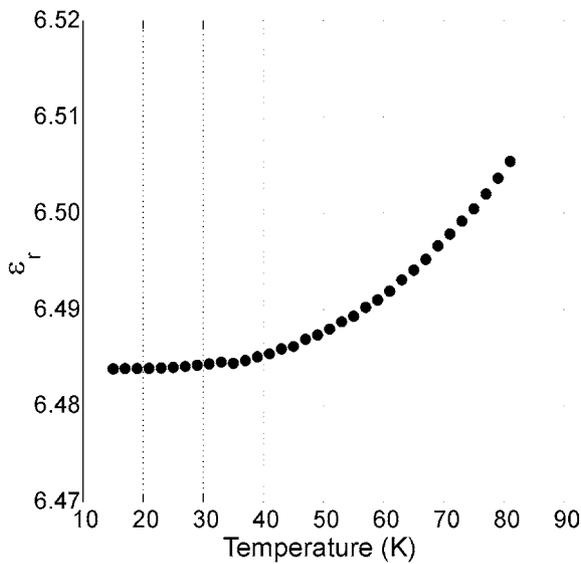

Fig. 5. Measured real part of permittivity of CaF$_2$ as a function of temperature at 29.25 GHz.

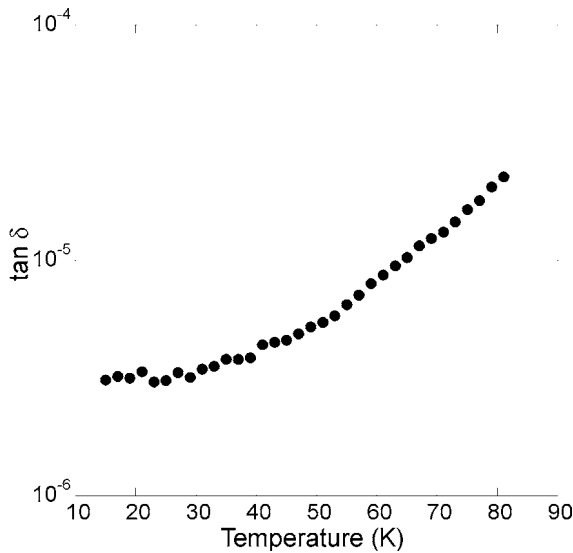

Fig. 6. Measured loss tangents of CaF$_2$ as a function of temperature at 29.25 GHz.

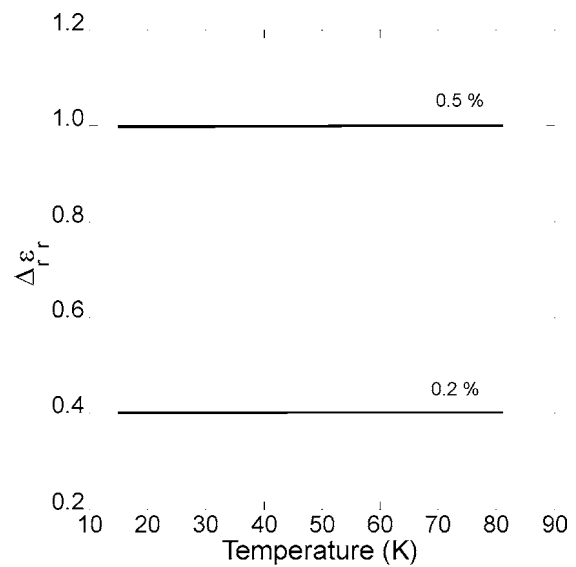

Fig. 7. Most probable error in $\varepsilon_r$ versus temperature for 0.2 and 0.5% uncertainty in samples' dimensions.

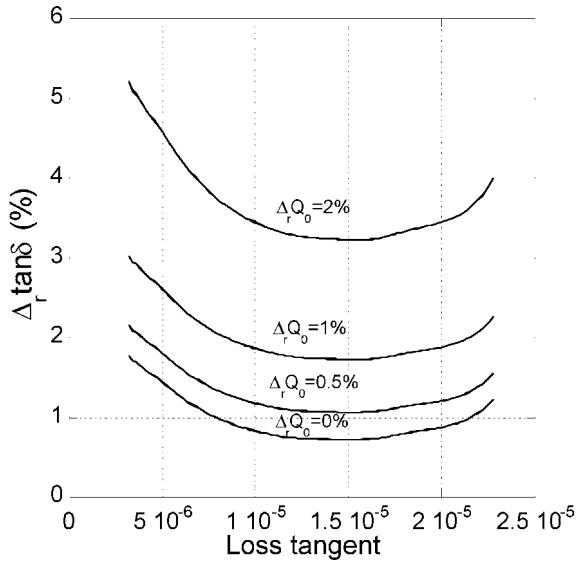

Fig. 8. Most probable error in tan$\delta$ of CaF$_2$ versus temperature for varying uncertainty in $Q_o$.

Table 2
Dimensions and permittivity and loss tangent of CaF$_2$ at 15 and 81 k

|  | Temp. (K) | CaF$_2$ |
|---|---|---|
| Height (mm) | 15 | 3.039±0.006 |
|  | 81 | 3.040±0.006 |
| Diameter (mm) | 15 | 4.971±0.009 |
|  | 81 | 4.973±0.009 |
| Permittivity | 15 | 6.483±0.026 |
|  | 81 | 6.505±0.026 |
| Tan$\delta$ ($\times 10^{-6}$) | 15 | 3.1±0.09 |
|  | 81 | 22.7±0.68 |
| Frequency (GHz) |  | 29.25 |

To assess errors in our measurements of tan$\delta$ we assumed uncertainties in $R_{SS}$ and $R_{SM}$ of 2%, and 0.5% for uncertainties in $A_S$, $A_M$ and $\rho_e$. Calculated errors in measured loss tangent values of CeF$_2$ for assumed uncertainties in the $Q_0$-factor measurements of 0, 0.5, 1 and 2% are presented in Fig. 8. The calculated MPE in tan$\delta$ for perfect $Q_0$-factor measurements is approximately 1%. We assess the uncertainty in the $Q_0$-factor of our measurement system as 1%. Hence the most probable error in tan$\delta$ of CaF$_2$ (which varied from $3.1\times10^{-6}$ to $22.7\times10^{-6}$) is between 1.7 and 3%.

## 5. Conclusions

The real relative permittivity and loss tangent of CaF$_2$ have been measured at frequency of 29.25 GHz at cryogenic temperatures using the Hakki-Coleman dielectric resonator with superconducting endplates. The recently developed Transmission Mode $Q$-Factor technique was used for data processing to remove noise, crosstalk and delay due to un-calibrated cable and connectors and to ensure high precision of measurements. Calcium fluoride was found to exhibit $\varepsilon_r$ varying from 6.483 to 6.505, and tan$\delta$ from $3.1\times10^{-6}$ to $22.7\times10^{-6}$ in the temperature range from 15 to 81 K as given in Table 2. On the basis of performed error analysis we assessed the uncertainty in the measurements of $\varepsilon_r$ and tan$\delta$ to be below 0.4 and 3%, respectively in the temperature range from 13 to 81 K. Our measurements have shown that CaF$_2$ is a very low loss material at cryogenic temperatures at frequency of 29 GHz. Hence, apart from optical applications, calcium fluoride can be useful in cryogenic microwave circuits where very low tan$\delta$ and resistance to atmospheric conditions are needed and the real relative permittivity of approximately 6.5 is adequate.

## Acknowledgements

This work is done under the financial support of ARC-Large grant (A00105170) James Cook University. The first author acknowledges the James Cook University Post Doctoral Fellowship and the MRG.